\documentclass[11pt]{article}
\pdfoutput=1
\usepackage[margin=1in]{geometry}
\usepackage{amsmath,amssymb,graphicx}
\usepackage{hyperref}
\usepackage{slashed}
\usepackage{verbatim}

\usepackage[T1]{fontenc}
\usepackage[utf8]{inputenc}

\usepackage{titlesec}
\renewcommand{\thesection}{\Roman{section}}
\titleformat{\section}[block]{\bfseries\centering}{\thesection.}{1em}{}
\renewcommand{\thesubsection}{\arabic{subsection}}
\titleformat{\subsection}[block]{\bfseries\centering}{\thesection.\thesubsection}{1em}{}

\usepackage[font=small,labelfont=bf]{caption}
\usepackage{cite}

\newcommand{\be}{\begin{equation}}
\newcommand{\ee}{\end{equation}}

\usepackage{xcolor}
\usepackage{bm}

\usepackage{lineno}

\title{\bf  Snowmass Theory Frontier \\
Effective Field Theory Topical Group Summary}

\newcommand\snowmass{\begin{center}\rule[-0.2in]{\hsize}{0.01in}\\\rule{\hsize}{0.01in}\\
\vskip 0.1in Submitted to the  Proceedings of the US Community Study\\ 
on the Future of Particle Physics (Snowmass 2021)\\ 
\rule{\hsize}{0.01in}\\\rule[+0.2in]{\hsize}{0.01in} \end{center}}

\begin{document}

\maketitle
\thispagestyle{empty}

\begin{center}
\author{Matthew Baumgart, Fady Bishara, Tomas Brauner, Joachim Brod, Giovanni Cabass, Timothy Cohen, Nathaniel Craig, Claudia de Rham, Patrick Draper$^{*}$, A. Liam Fitzpatrick, Martin Gorbahn, Sean Hartnoll, Mikhail Ivanov, Pavel Kovtun, Sandipan Kundu, Matthew Lewandowski, Hong Liu, Xiaochuan Lu, Mark Mezei, Mehrdad Mirbabayi,  Ulserik Moldanazarova, Alberto Nicolis, Riccardo Penco, Walter Goldberger,  Matthew Reece,  Nicholas L. Rodd, Ira Rothstein$^{\dagger}$, Shu-Heng Shao, Will Shepherd, Marko Simonovic, Mikhail Solon, Dam Thanh Son, Robert Szafron, Andrew Tolley, Zhengkang Zhang, Shuang-Yong Zhou, Jure Zupan}
\end{center}


\bigskip

\begin{center}
$^*$pdraper@illinois.edu,
$^\dagger$izr@andrew.cmu.edu
\end{center}

\bigskip

\snowmass


\newpage
\pagenumbering{arabic}

\section{Introduction and Executive Summary}

Quantum field theory is the fundamental mathematical structure  describing the world.
Not surprisingly,  its power and generality leads to overwhelming mathematical challenges in its application to the
real world, i.e., in non-idealized situations.  A primary source of complexity arises from the fact that
typical physical systems depend upon physics arising from hierarchically distinct length scales.
However, given the local nature of physical laws, we may disentangle these scales in a way that reduce the computational complexity for any given prediction. This procedure involves finding the proper action to capture the physics at a particular length scale. 
These actions are  constrained by the  symmetry breaking pattern which
fixes the number of unknown parameters that either need to be fit from experiment or calculated using the knowledge of the short distance theory.
Finding a systematic expansion
within which to design these actions is the art of effective field theory (EFT).

 EFTs were first utilized in a modern sense in describing  pion physics and later the weak interactions, though much of the    physical insight, including the use  of  the renormalization group, came from  the realm of critical phenomena.  While the EFT approach may be said to have been championed by the HEP community, it is has been utilized to great success in many fields and sub-fields of physics, as described in various contributions to these proceedings. Here we summarize the recent process and future opportunities reviewed in detail in the whitepapers~\cite{Cabass:2022avo,deRham:2022hpx,Baumgart:2022vwr,Craig:2022uua,Cohen:2022tir,Shepherd:2022rsg,Brauner:2022rvf,goldberger}.

Dark matter detection relies on EFTs for both direct and indirect measurements in various contexts, including systematizing the low energy interactions with nuclei and re-summing large logs in cosmic
annihilation processes~\cite{Baumgart:2022vwr}. In cosmology, EFTs have been  utilized to great success to study Large Scale Structure (LSS) and inflation~\cite{Cabass:2022avo}. In particular, EFT calculations have led to state-of-the-art predictions for cosmological parameters
from the LSS data that compete with those from the cosmic microwave background.  In models of inflation, EFT techniques have been used to place
novel bounds on non-Gaussianities.

The ubiquitous nature of EFTs has lead to inter-disciplinary cross pollination. 
Condensed Matter Physics  (CMT) is another natural setting for EFT methods, where they  has been utilized to shed light on various systems, including out of equilibrium phenomena, hydrodynamics, and (non) Fermi liquids~\cite{Brauner:2022rvf}.
Recently there has been  a strong inter-disciplinary movement leveraging EFTs to understand
new exotic states of matter called ``fractons." EFTs have also played an important role in the
nascent field of gravitational wave astronomy, with the development of an EFT to  
predict the signatures of binary inspirals~\cite{goldberger}.  Scattering amplitudes techniques can be used within this EFT to streamline-higher order calculations, much like in the case of QCD, and using these ideas, state-of-the-art calculations have
been performed by people within the HEP community.  Given the relative newness of this field and
the wealth of new data, EFTs will play an important role in future theoretical efforts.

EFTs also continue to play a central role in accelerator physics. Soft-Collinear Effective Theory (SCET)
has greatly improved our understand of factorization in high energy scattering, which is a necessary
ingredient for any theoretical prediction in order to disentangle the physics of the proton from 
the high energy scattering process of interest. Moreover, SCET is utilized to sum large logarithms
which lead to poorly-behaved perturbative series. Its value has lead SCET to become its own sub-field, with dedicated yearly conferences. 

At the LHC, a core component of the search for new physics involves  treating the Standard Model as an effective theory  (SMEFT) and looking for signatures associated with its higher dimensional operators.
There has been considerable progress in constraining the values of the Wilson coefficients
which capture the unknown UV physics of interest and systematizing calculational procedures~\cite{Shepherd:2022rsg,Cohen:2022tir}. The SMEFT formalism will continue to be a staple of experimental analyses.

The examples above highlight the critical role of EFTs  in the direct comparison of theory with observation across a huge range of physical systems. Thinking critically about the self-consistency of our EFT descriptions of nature has also played an essential role in establishing the ``big" questions that physics should seek to answer. The major naturalness puzzles -- the electroweak hierarchy problem, the strong CP problem, and the cosmological constant problem -- are strongly suggestive of new dynamics and principles, and they continue to drive creative new theoretical approaches and experimental designs in tandem~\cite{Craig:2022uua}. At the same time, more formal studies of EFT have resulted in  new methods to constrain consistent low energy theories~\cite{deRham:2022hpx,Draper:2022pvk}, exploiting basic principles both of QFT and of quantum gravity. These approaches have lead to  testable predictions about the possible forms new physics can take.

EFTs have become an essential tool in many areas of physics and are continually being developed.
Given that  the scientific output of any experiment is bounded by the accuracy of the theoretical
predictions, it is not an understatement to say that EFTs play a vital role in any scientific program.
The HEP community has leaned heavily on this formalism and the need for further developments
in this field is driven by experimental exigencies. 
Meanwhile more formal EFT research continues to provide new tools for attacking some of the deepest questions in the field.

\label{sec:intro}


\section{Dark Matter}
\label{sec:darkmatter}

It is difficult to overstate the scientific importance of the dark matter problem.
This is evidenced by the tremendous amount of resources presently being focussed to find the solution.
The list of experiments, both large and small, dedicated to  this problem is an increasing function of time.
The  reason for the challenging nature of   DM detection is not difficult to discern given its name.
The design of an appropriate experiment involves some speculation as to  the
dark matter abundance and interaction rate, and this relies heavily on  the theoretical input
which quantifies the reach of an experiment in the space of DM parameters.

EFT has played in increasingly important role in both direct and indirect dark matter detection~\cite{Baumgart:2022vwr}.
In case of the former,  for heavy DM candidates ($M\gg M_W$), the cross section for DM  to interact with nuclei can be described a heavy particle effective theory which is derivative of the heavy-quark effective theory (HQET). In the heavy mass limit spin-independent WIMP-nucleon scattering cross sections become universal for given
WIMP electroweak quantum numbers. Moreover, the EFT analysis revealed that the generic amplitude involved cancellations which suppress the cross section orders of magnitude below naive estimates.
Furthermore, working within the EFT, one is able to systematically include the effects of QCD corrections. When the mediator mass is larger than the momentum exchanges one can then write down
as set of operators where by the DM has contact interactions with nucleons, leading to
a reduced operator basis which can be scanned by experiment.

The nature of the relevant EFT changes when considering indirect detection, as now the 
physics involves a semi-inclusive annihilation process. 
If the DM is charged under $SU(2)_L \otimes U(1)_Y$
(or another force with light mediators), then it is subject to a long-range potential  (when $M_{DM}\gg m_W$) which can boost its annihilation rate via a so-called  ``Sommerfeld enhancement". Moreover,  when calculating the gamma ray spectrum from  annihilation, the  hierarchy of scales  leads to large
(double) logarithms that may necessitate resummation. Thus the  EFT in this case is richer
as it must be able to systematically describe both the Coulombic infra-red singularities as well
as the soft and collinear singularities which arise due to gauge boson emission.
Since the DM velocity is smaller for the signal events then in the early universe freeze-out, 
Coulombic corrections  are more enhanced  in the former case.
 
 To resum the Coulombically enhanced terms, which form a power series in the ratio $m_{DM}/m_W$, one may lean upon  NRQCD (Non-Relativistic QCD), the EFT developed to describe quarkonia in a systematic
expansion in the relative velocity.  The Next-to-Leading order corrections  from these resummations can be a large as twenty percent.
The singularities in the radiation can be calculated by utilizing Soft-Collinear Effective Theory (SCET),  developed to describe high energy scattering.  By expanding around null eikonal lines it is possible
to use this theory to resum the aforementioned IR logs using standard renormalization group techniques. In addition,   when looking for line spectra one must account for the finite energy 
resolution of the detection, as hard radiation (from $2 \rightarrow 3$  processes) near the end-point will begin to contribute.
SCET can be used to resum end-point logs of the form $\log(1-z)$,
where $z$ is the energy fraction below which the photon could no longer have been a product of
a $2 \rightarrow 2$ process.

 Outside of heavy WIMP scenarios EFT's have been utilized  in detecting sub-GeV DM candidates 
where nuclear  recoil is highly suppressed and no longer viable.
Instead, one may take advantage  interactions between the DM and the low lying modes in condensed matter systems, either emergent ones such as  phonons or magnons, or via scattering off of electrons. The EFT for emergent modes can be built in a systematic fashion following standard space-time symmetry breaking coset constructions (see the next section).
Publically available code which implements the EFT framework has been produced. 
Using  EFT's  to detect dark matter via table top  condensed matter experiments is an active field
which has gathered much attention in both the theoretical and experimental community. Given the
rapidly evolving detector technology this direction holds considerable promise to probe previously
unreachable regions of the dark matter parameter space.


\section{Applications to Cosmology}
\label{sec:inflation}

EFTs have played an  increasingly important role in improving the accuracy of theoretical predictions during this era of precision cosmology.  They have been applied primarily to three areas of cosmology: Inflation, Large Scale Structure (LSS), and Dark Energy (DE)~\cite{Cabass:2022avo}.

  Inflation necessitates  some light  dynamical degree of freedom in order for the inflationary  epoch to end.
This degree of freedom plays the role of a ``clock"  that tracks the transition from the accelerating to de-accelerating phase. The clock field acts as an order parameter which  breaks time translation invariance and its fluctuations represent the relevant Goldstone boson ($\pi$) which non-linearly realizes the broken symmetry. Furthermore, it can be shown that the fluctuations in the Goldstone field are, in fact, related the curvature fluctuations ($\xi$) 
such that the $\xi$ correlation function  may be calculated in terms of those of $\pi$.
The primary inflationary observables of interest are the power spectra and deviations from Gaussianity (via study of the bispectrum) which are sensitive to the model dependent interactions.
These observables can be calculated via the EFT for $\pi$ in a systematic expansion in derivatives.  More complicated models involving multiple clock fields have also been proposed and there exists a set of constraints on single clock inflation which can be used
to distinguish between the two classes of theories.

The search for   non-Gaussianities is performed on the Cosmic Microwave Background (CMB) as 
well as LSS, which is defined as the matter density distribution at large red-shifts.
Searching for   non-Gaussianities   via LSS  has the advantage
that  the number of modes available for measurements  is  much larger than that of the CMB because the matter distribution is essentially three-dimensional.  The EFT of LSS is rather unique in several ways. Firstly, it is formulated at the level of the (fluid) equations of motion.
Furthermore the correlators are statistical  (non-thermal) averages. Nonetheless all of the 
usual technology of EFT and QFT, i.e. Feynman diagrams, loops etc. can be utilized.
Loop corrections have been calculated up to three loops for the power spectrum and two loops
in the bispectrum.

EFT analyses have led to     bounds on the dark energy equation of state as well  as precision extractions of various cosmological parameters including the 
Hubble constant, the power spectrum amplitude and tilt. These measurements compete with the bounds derived  from the CMB.
More recently, the EFT of LSS has been used to place
novel  bounds on single-field primordial non-Gaussianity from galaxy surveys. With the wealth of new data expected, precision predictions from LSS EFT will be a vibrant sub-field.

EFTs are utilized to look for variations of  Einstein-Hilbert  gravity. However, GR  is a  parsimonious  and elegant theory. It is the unique theory of a spin two massless particle, 
and it resists tampering, though  there is motivation to  do so, as the minuteness of the
cosmological constant is relentlessly taunting.  Thus there has been significant effort in the theory
community to try to modify gravity at long distances. 
This can be done by either explicitly adding light degrees of freedom or by  explicitly breaking diffeomorphism 
invariance. 
 There are no theoretical obstructions to adding scalar degrees of freedom, but upon doing so one is immediately faced with  stringent  bounds on long range forces especially from, solar system bounds.
Breaking diffeomorphism invariance on the other leads to other complication including the generation of 
a strong coupling scale. The challenge being taken up by the community is to find an EFT of
GR which can explain the acceleration at cosmological distance scales while avoiding all of
the constraints from shorter distance scale physics. 


\section{EFT and UV Principles}
\label{sec:UVprinciples}

The standard principle used in constructing EFTs is to enumerate the light degrees of freedom and write an effective Lagrangian including all effective operators consistent with any exact symmetries. The Wilson coefficients are arbitrary, but in the absence of approximate symmetries they are generally taken to be order one in units of the cutoff. This principle has been used to great effect, and it can be checked in any number of examples. 

At this level the space of EFTs appears vast. However, there is now considerable evidence that  the true space of EFTs is subject to a panoply of intricate and stringent constraints imposed by the existence of a consistent UV completion. The fundamental properties of causality and unitarity have been used to great effect in deriving constraints on Wilson coefficients in both nongravitational and gravitational EFTs, while closely related methods have been used to map general properties of CFTs with holographic duals to gravitational EFTs in the AdS bulk. Complementary to this, the swampland program has established  evidence for conjectures about the general properties of EFTs that can be  embedded in theories of quantum gravity.

In recent years there has been enormous progress in developing these constraints, applying them to map the space of consistent EFTs, and exploring phenomenological implications. Summarizing the WP~\cite{deRham:2022hpx}, and also drawing on the WP~\cite{Draper:2022pvk}, this section will review recent progress and future opportunities in the vigorous theoretical research program exploring the ways in which EFTs can be constrained by principles of ultraviolet consistency.

One of the most rigorous ways to map the space of consistent EFTs is to study the consequences of requiring the UV completion itself to be a consistent QFT, respecting fundamental properties of causality and unitarity. Causality implies analyticity properties of correlation functions as a function of complexified coordinates, which can be applied to particular effect in CFTs, resulting in nontrivial constraints on the light operator spectrum. In more general QFTs it is more convenient to work with the S-matrix. 

Unitarity bounds on the growth of scattering amplitudes with Mandelstam invariants, together analyticity properties following from causality, allow the construction of dispersion relations satisfied by amplitudes in consistent QFTs. These dispersion relations, in turn, supply constraints on the derivative expansions of amplitudes, typically in the form of positivity bounds. These are low energy bounds that can be interpreted directly as constraints on the signs of Wilson coefficients in consistent EFTs. There has been great progress in developing this program in recent years, resulting in infinite numbers of bounds; bounds on theories with massive particles of arbitrary spins; the discovery of positivity bounds on nonlinear functions of EFT couplings; and the construction of lower and upper bounds on Wilson coefficients by combining nonlinear bounds with crossing symmetry constraints. The latter leads to finite "islands" in parameters pace.  These are closely related to the S-matrix bootstrap~\cite{Kruczenski:2022lot}. 

Similar methods can be used in gravitational EFTs, in which a massless spin-2 graviton is part of the low-energy spectrum. The graviton leads to some complications, but progress in recent years lead to the  first bounds on the leading corrections to general relativity.

In a complementary direction, it has long been known that causality constraints on Wilson coefficients can sometimes be seen directly in the EFT by examining signal propagation on nontrivial backgrounds. This approach is particularly direct in ordinary QFTs, but is challenging in gravitational EFTs. Recent progress has begun to surmount this problem, leading to the development of "infrared causality" criteria based on the scattering time delays experienced by long and short wavelength modes. These methods complement the S-matrix positivity bounds and can be used to constrain EFTs in some cases where the S-matrix does not exist. 

A third approach uses CFT data to constrain gravitational EFTs in AdS with holographic duals. The general properties of CFTs with bulk duals strongly constrain the form of CFT correlation functions involving the stress tensor, which can be translated into properties of the bulk gravitational EFT.   In some cases, the conformal bootstrap~\cite{Hartman:2022zik,Poland:2022qrs} can be used to rigorously establish bounds in asymptotically AdS  theories directly analogous to S-matrix bounds in asymptotically flat space which rely on well-motivated but unproven assumptions. Even small, $O(1/M_p^2)$ violations of strict positivity bounds known to arise in flat space theories with dynamical gravity have been founds also in the AdS/CFT context.

The EFT bounds described thus far descend from general principles, with little input from the gravitational sector of the UV completion. What is now generally known as the Swampland program seeks to infer the universal properties of EFTs that can be consistently embedded in a theory of quantum gravity. Since a general definition for quantum gravity and the universal properties shared by all consistent theories are not rigorously established, these properties are conjectural. However, in many cases they are supported by a diverse spectrum of evidence from perturbative string theory, semiclassical gravity, AdS/CFT, and other sources, and a significant part of the program is devoted to mining new evidence and refining the conjectures. In many cases the swampland arguments lead to entirely new constraints on EFTs, while in some cases they can be directly related to positivity bounds of the types described above. 

Well-known examples of these sorts of criteria include the conjecture that theories of quantum gravity do not possess exact global symmetries, and the weak gravity conjecture (WGC), which constrains the spectrum of charged sources in gauge theories. Recently there has been great progress in refining and strengthening the WGC, and in attempting to determine how badly global symmetries need to be broken in low energy EFTs. In the ``no global symmetries" case, a diversity of arguments has lead to an emerging picture that the ``price" of an approximate symmetry's quality is tied to the UV cutoff on local quantum field theory: if the largest symmetry violating coupling is of order $c$, the cutoff scale is generally of order $\Lambda^2 \lesssim -\frac{1}{8\pi^2} M_p^2 \log(c).$ In the WGC case, it has been recognized that all examples in string theory in fact obey much stronger bounds than the original WGC, possessing infinite towers of superextremal particles of different charges. This has been elevated to a principle in the ``Tower WGC." These towers in fact satisfy another swampland conjecture, the distance conjecture (SDC), which predicts that an infinite tower of light states emerges in the asymptotic regimes of moduli space, which are also weak coupling limits in string theory. The towers also radiatively affect the strength of gravity, driving it to strong coupling at $\Lambda \ll M_p$. Thus these conjectures and their precise realizations are all tightly intertwined by addressing the same limit point where an exact global symmetry emerges. It is remarkable that global symmetries play a foundational role in the ordinary construction of an EFT, while the strength of their breaking or gauging is bounded from below in quantum gravity. 

An essential part of the swampland program is in elucidating phenomenological implications of the conjectures, surveyed in the WP~\cite{Draper:2022pvk}. Ideas like the WGC and its extensions, the ``no global symmetries" principle and more generally the cost of implementing approximate global symmetries, the SDC, the emergent string conjecture, and others have been shown to invalidate or impose strong constraints on many classes of models of new physics, while at the same time giving new motivation for other model-building ideas and suggesting interesting new ways that EFTs can break down. For one example, the WGC has been used to provide a powerful argument that the photon must be truly massless, while providing significant constraints on theories of dark photon dark matter. In another direction, the SDC and WGC both provide strong limitations on models of large field inflation, including solutions to the cosmological constant and electroweak hierarchy problems that invoke very lengthy periods of inflation. 
The no global symmetries principle  has also led to a recent reevaluation of solutions to the strong CP problem: on one hand, there are some hints that an axion with the right properties to relax the QCD $\theta$ term may actually be necessary in order to remove a generalized  symmetry, while on the other hand it can be viewed as providing new motivation for solutions to strong CP that do not involve such high quality approximate symmetries as the Peccei Quinn mechanism. These applications, as well as the recent bloom of conjectures and phenomenological studies related to the existence and stability of de Sitter vacua, are representative of the kinds of novel insights that quantum gravity can provide to  effective field theories of particle physics~\cite{Draper:2022pvk}.

There are also intriguing connections between the swampland conjectures and the positivity bounds and holographic approaches discussed above. At large charge the Tower WGC appears to imply the  existence of superextremal classical black holes, which can be rephrased as the existence of higher dimension operators in the gravitational effective action with certain positivity properties satisfied by their coefficients. Some of these positivity properties overlap with known constraints described above, while others involving curvature operators are are more subtle and under active investigation. Meanwhile holography has been used to prove the ``no global symmetries" conjecture for all  theories of quantum gravity in asymptotically AdS space, and there is considerable ongoing activity in translating the various WGCs in AdS into statements about dual CFTs, which might be tested with bootstrap or other methods.

The recent work described in this section has vastly improved our understanding of the fundamental principles satisfied by EFTs, with and without gravity. Looking to the future, there are many clear directions for development and discovery~\cite{deRham:2022hpx,Draper:2022pvk}. Examples include linking AdS/CFT methods to the S-matrix bootstrap by taking the flat space limit; extending unitarity and analyticity constraints to derive new positivity bounds from higher-point scattering amplitudes;  fully exploring the application of positivity bounds to the SMEFT; developing analogs of analyticity bounds for cosmological EFTs; connecting positivity and causality properties to the WGC for extended objects charged under generalized symmetries;  proving swampland conjectures via positivity or bootstrap methods; exploring how proposals for holography in de Sitter space support or refute the de Sitter swampland conjectures, and how they might impact inflationary observables; searching for general arguments in support of the emergent string conjecture or stringy examples that violate it, and studying its phenomenological applications; and further exploring possible roles for the WGC and the no global symmetries principles in explaining naturalness puzzles. These and other opportunities will source a vibrant research program with connections and impact on many other areas of high energy physics.



\section{Naturalness}

As mentioned in the previous section, a standard principle of EFT construction is to write all operators consistent with the symmetries with O(1) coupling strength. Famously, the SM appears to fail this principle: certain parameters are known empirically to be very small and yet they are unconstrained by any symmetry of the SM itself. The most striking examples are the strong CP problem, or why the CP-violating phase parameter $\theta$ of QCD is smaller than $10^{-10}$; the electroweak hierarchy problem, or why the weak scale $m_{W,Z,h}\sim 100{\rm GeV}$ is so far below UV scales associated with quantum gravity, unification, and other physics beyond the SM; and the cosmological constant problem, or why the vacuum energy density $\rho_{vac}\sim (10^{-3}{\rm eV})^4$ is so much smaller than nearly every other physical scale observed in the SM or believed to lie beyond it. These ``naturalness problems" may signal the presence of new physics beyond the SM, with new symmetry structures or dynamics to explain the small numbers and scale hierarchies within a field theory framework. Alternatively they may signal that the basic EFT principle breaks down for some reason. Either possibility has enormous significance for our understanding of fundamental physics. 

Naturalness problems are of fundamental importance  and have a lengthy history. These problems have been with us for decades, stimulating vast bodies of literature and experiments. We have learned much about nature and about field theories from this work. Thus far, we still do not know the resolutions to these problems, and so they continue to represent essential goals for the field. Moreover, they continue to stimulate new ideas: recent years have seen continued progress in model building along creative new axes and an explosion of progress in thinking about new experimental signatures. Here we give a brief and general overview of some of these developments for each of the three major naturalness problems, based on the WP~\cite{Craig:2022uua}.

Broadly speaking, there are two known ways to address the strong CP problem. The first is to use a combination of symmetry and dynamics to set $\theta=0$ as a UV boundary condition on the standard model. Within the SM itself the renormalization of theta is tiny. The second method is to relax theta in the infrared with an axion. These two general approaches have been known for many years, but both have been revitalized in the theory community in the last several years.

In recent years there have been a number of phenomenological profiles and model-building developments in spontaneous P/CP breaking UV solutions to the strong CP problem. These studies have revealed that instead of living inaccessibly far in the UV, many of the most successful implementations of these mechanisms predict a spectrum of collider, EDM, and cosmological signatures accessible in near-term experiments. Both the signature space and model space are still being charted and offer more opportunities for theoretical progress. Some aspects were reviewed in the WP~\cite{Blinov:2022tfy}.

Axion solutions to the strong CP have undergone a renaissance since the last Snowmass, most visibly with the explosion of progress in mapping out the astrophysical, cosmological, and terrestrial signatures of axion dark matter, and opportunities for refinement of these ideas and exploring new potential signatures abound. The model-building side has seen new proofs-of-concept for populating the essentially the entirety of axion-like-particle mass and coupling space. WPs surveying the state of axion theory and experiment include~\cite{Blinov:2022tfy,Adams:2022pbo,Agrawal:2022yvu}.

Curiously, the proposed solutions to strong CP can sometimes introduce new naturalness problems as severe as the one they are introduced to solve. With UV solutions the most notable example relates to the stability of the scale of spontaneous P/CP violation, and recent works have broken new ground  embedding these solutions in natural UV completions. For the axion, the famous "quality problem" refers to the fact that the mechanism requires a global shift symmetry that is only broken by QCD to a part in $10^{10}$. Here a number of recent papers have investigated axions in string theory and found indications that in fact the quality problem is not so severe, and furthermore that an axion with the right properties may in fact be an essential feature of quantum gravity related to the absence of generalized global symmetries. Future theoretical investigations in these directions are likely to lead to new understanding of what it means for a solution to strong CP to itself be natural and refinement of space of well-motivated models.

The cosmological constant (cc) problem is the most numerically severe naturalness problem. In field theory the vacuum energy density appears to receive contributions from many and varied sources, including SM particle thresholds, the chiral condensate, the tree level Higgs potential, and any new ultraviolet degrees of freedom. All of these point to a vacuum energy vastly larger than what is observed, indicating either enormous cancellation, a dynamical mechanism, or a subtle reason why all of these field theory estimates are incorrect. 

Among the dynamical mechanisms, the idea that a relaxation process might automatically select a small cc has recently been revived and developed in interesting ways. The general idea of relaxation is quite old, but was abandoned for many years because it was thought to predict a big but completely empty universe, throwing out the baby with the bathwater. Recent progress has surmounted this long-standing problem, showing that it is possible, at least in principle, to construct models that both relax the cc and successfully reheat. This efforts have the potential to develop into fully realistic relaxation solutions to the cc problem.

Another active area of research concerns the possibility that there is something wrong a priori with the effective field theory computations of corrections to the cc. Particular scrutiny has fallen on the possibility that ``UV/IR mixing" effects associated with quantum gravity might cause EFT to badly overestimate these contributions. 
It has been suggested by a number of authors for many years that the holographic principle might play a role in addressing the cc problem, and recent work has focused on the Cohen-Kaplan-Nelson bound, a mysterious and rather dramatic correlation between UV and IR scales in a field theory coupled to gravity. The meaning of this correlation is as of yet not well understood, but recent work has given new interpretations in terms of thinning the independent degrees of freedom, which is at least able to eliminate the quantum contribution to the cc from UV modes. At the same time many recent phenomenological studies have lead to new ideas and directions for exploring the possible experimental implications of UV/IR correlations. This represents a general growth area both in the study of the cc and in the ways that quantum gravity might impact low energy EFTs, as also discussed above in previous sections.

Like the cc problem, the electroweak hierarchy problem results from the strong sensitivity of superrenormalizable parameters in generic quantum field theories to ultraviolet physics. In this case the relevant parameter is the SM Higgs mass parameter, which receives quantum corrections quadratic in UV scales. The seesaw mechanism, grand unification, the inflationary sector, and innumerable other UV extensions or completions of the SM generically provide gigantic additive corrections to the Higgs mass. 

Conventional dynamical resolutions to the EWHP include technicolor, low-scale supersymmetry, and Higgs compositeness. The discovery at the LHC of a relatively light Higgs boson, as well as the non-observation of any new particles or deviations in Higgs couplings from SM predictions, has excluded or placed stringent bounds on each of these ideas. Although it continues to be useful and interesting to explore implementations of these basic mechanisms that might avoid the constraints, the pressure on them has resulted in a wealth of new ideas for novel mechanisms and their experimental signatures. To pick an example, one area undergoing the most active development in the last few years has been the neutral naturalness paradigms. 

Neutral naturalness models extend the SM so that the Higgs mass is protected by a discrete symmetry. They derive their name from the property that none of the new states responsible for the protection of the weak scale need to be charged under the SM gauge groups. As a result they present an entirely new way of thinking about naturalness experimentally. Unlike the spectacular high-energy signatures of new charged and colored states associated with technicolor and supersymmetry, the signatures of neutral naturalness tend to be more closely tied to the Higgs, including Higgs decays into long-lived light particles and Higgs coupling deviations. Models and signatures are reviewed in the dedicated WP~\cite{Batell:2022pzc}. There are also cosmological signatures connected with the partners of the light SM states under the discrete symmetries, reviewed in the WP~\cite{Dvorkin:2022jyg}.

Other recent and novel lines of attack on the EWHP include new models of cosmological relaxation; models with large numbers of hidden copies of the SM, so that a low electroweak scale emerges in one of them statistically; ideas tying the weak scale to the weak gravity and other swampland conjectures; applications of self-organized criticality; and the possibility that UV/IR mixing effects might again play a role in altering EFT reasoning. All of these represent creative and promising directions for theoretical development. In many cases there are intriguing "proofs of principle," hinting at the possibility of much larger yet-to-be-discovered spaces of models and mechanisms.

Some general lessons can be drawn from much of the recent work on each of the naturalness problems. First, experimental signs of naturalness are much broader than previously imagined. This has been recognized through vigorous phenomenological efforts to invent and explore both new models and new observational techniques. Second, basic theoretical mechanisms can often be repurposed from one naturalness problem to another, and sometimes a solution to one problem impacts aspects of other problems. Thus it can be beneficial to view these problems holistically. Third, new developments in formal theory topics, including the amplitudes, generalized symmetry, swampland, and string phenomenology programs, can play an important role in stimulating new ideas for model-building. Casting a broad net, both theoretically and experimentally, will be important to ensure that the role of EFT in nature, and its limits, are fully understood.

\section{SMEFT}
\label{sec:SMEFT}
The Standard Model EFT (SMEFT) extends the perturbatively renormalizable Standard Model to include higher-dimension operators with coefficients suppressed by inverse powers of a characteristic new physics (NP) scale $\Lambda$, generally of order the lightest particle mass in the NP sector. As such it provides a largely model-independent framework to parametrize the effects of NP on SM scattering processes with energy below $\Lambda$. The power of the SMEFT is that one can compute a large library of collider observables in the SMEFT to a specified and systematically improvable precision, fit or constrain the Wilson coefficients from data, and finally compare with any heavy NP model by matching the latter onto the SMEFT. This procedure factorizes the experimental and observable components of the problem, which can be handled once and for all inside of the SMEFT, from the more abstract (and typically much simpler, now substantially automated) process of integrating out NP, which can be done on a model-by-model basis. As carrying out precision computations of collider observables within every NP model of interest is intractable, an EFT framework like the SMEFT is in fact essential.

Significant progress has been made since the last Snowmass in extensively developing the SMEFT methodology and applications, reviewed in the WP~\cite{Shepherd:2022rsg}. On a technical level, progress includes extensions to higher orders in the EFT expansion, higher-precision matching methods, development of computational tools, map making between different EFT frameworks, and establishing well-motivated approaches to uncertainty estimation. Many of the applications have fallen into two broad classes: (1) to processes with characteristic energies around the electroweak scale $E\sim v$, typically with high statistics observables involving the on-shell production and decay of SM states, and (2) to high energy $E\gg v$ tails of distributions, where statistics are lower but the effects of higher-dimension operators $\sim E/\Lambda$ are enhanced. In the former class, it has been shown that a finite number of combinations of Wilson coefficients can contribute to processes with sufficiently restricted kinematics, like the on-shell production and 2-body decay of a SM particle. These combinations have been efficiently repacked in a "geometric formulation" of the SMEFT, and they allow the finite parametrization of heavy NP effects on the process to all orders in the EFT expansion. For processes with more particles or some off-shell states, the parametrization is no longer finite, but it can be made so by focusing on the kinematic regimes where all invariants are of order $v$ and higher derivative effects can be neglected to high accuracy. In the case of high-energy tails, progress has been made in establishing rigorous treatments of the perturbative expansion in scale hierarchies and in estimating the theoretical uncertainty induced by truncating the SMEFT at finite operator dimension. These studies are essential to making as-accurate-as-possible maps from a truncated SMEFT to observables at high invariant mass while giving a balanced estimate of the effects of uncertainties on fits. At high energies a careful treatment of theoretical uncertainties in fitting the SMEFT to data is particularly crucial, because the effects of missing higher-dimension operators are growing and must be estimated carefully against the growing effects of the higher-dimension operators that are included in the calculation.

Via the SMEFT program, the theory community has constructed a valuable framework for maximizing the theoretical implications of data from the long-term LHC program and future colliders.  The WP~\cite{Shepherd:2022rsg} identifies several essential areas for future development, including: continued research into SMEFT predictions at higher orders;
community adoption and refinement of uncertainty estimation algorithms that appropriately weight the possible effects of missing higher-order terms in the EFT expansion; 
development of new fitting algorithms that incorporate the parameter-dependent uncertainties;
production of sequential fits and likelihood functions; and
moving toward a global fitting framework that treats SM and EFT parameters on an equal footing. 
These goals provide a roadmap toward the ultimate target of building a reliable and versatile tool, capable of 
utilizing all information in 
precision SM measurements to 
fully explore the implications for high energy physics beyond the standard model.

We close this section with a brief summary of progress in matching, reported in the WP~\cite{Cohen:2022tir}. SMEFT and similar EFTs factorize the specifics of new physics from the comparison of theory with data. Given a precision SMEFT fit, to compare a particular BSM model with data and constrain or fit its parameters, one must first match the model onto the SMEFT. In recent years this procedure has been systematized to high precision, streamlined, and  automated. The WP~\cite{Cohen:2022tir} summarizes the ``functional matching" approach that has facilitated much of this progress. Functional matching affords an elegant, simple prescription for computing the effective Lagrangian obtained by integrating out heavy degrees of freedom from a UV theory at one loop order. In essence, it extracts the UV part of functional determinants as a function of the light fields, automatically providing a local EFT expansion for the latter via a covariant derivative expansion that explicit maintains the underlying symmetries. These methods have been widely applied in matching numerous UV models to the SMEFT and other EFTs. Furthermore, a number of codes have been written to implement functional matching and output EFT coefficients. Important future targets for developing the matching toolkit include interfacing codes directly with SMEFT fit codes; extending the functional techniques to other classes of EFTs like SCET; and going beyond one-loop order.


\section{Condensed matter}
\label{sec:CM}

The cross fertilization of condensed matter and high energy physics follows naturally
from the fact that they both have 
effective  field theory \footnote{While this is the unique description for HEP, the same can not be said for CM.} descriptions~\cite{Brauner:2022rvf}
.
Going back to Anderson's ideas regarding the lack of a Goldstone boson in systems with
spontaneously broken symmetries,  each field has consistently benefitted from insights generated from the other. The strong overlap is a consequence of the fact that, almost by definition,  continuum field theoretic 
descriptions  of condensed mattter systems are effective field theories, unless they are sitting exactly at the critical point.
Thus, EFT methods developed for HEP are sharp tools that can naturally adapted in study condensed matter systems.

HEP studies  typically assume Poincare invariant field theories as a starting point whereas by definition condensed matter systems break space time symmetries.  The absence of  linearly realized Lorentz invariance  in theories of CM leads to more complex actions which nonetheless are still under systematic control.
Given the symmetry breaking pattern, one may construct the EFT using various techniques to ensure that the broken symmetries are non-linearly realized. 
Using such techniques there has been significant progress in making predictions  for the dynamics of systems involving both collective and electronic degrees of freedom.  Condensed matter systems may have additional emergent  linearly realized symmetries that can lead to new exotic behavior such
as fractons (to be discussed below) and fractional quantum Hall states.

 EFTs have been utilized in new ways to describe systems out of equilibrium via the closed time path formalism where all degrees of freedom are doubled  and   the breaking of time reversal invariance and  dissipation are allowed. Such field theories have yielded  new descriptions of hydrodynamics even in cases where conventional descriptions breakdown.  The EFT approach to hydrodynamics has shed light on the chiral magnetic effect,  in classifying transport coefficients and applied the constraints of generalized higher form symmetries to the study of magneto-hydrodynamics.
 More generally,  EFT description of non-equilibrium systems have been utilized to describe a wide variety of systems including, liquid crystals, fluids with chemical reactions, black hole mergers as well as many others.

In recent years there have been proposals for various new phases of exotic matter which fall
under the rubric of ``fractons" which are defects with limited mobility.  Some fractons can not move
in isolation while other can only move along submanifolds. These models models are usually defined on the lattice and finding their proper EFT (continuum) description  and in some  cases has turned out be quite a challenge.
In a class of models taking the continuum limit leads to an infinite set of charges which implies extreme
sensitivity of the IR to the UV. There are many open questions as to how to find the proper continuum limit (EFT) for fractonic theories, as they seem to fly in the face of accepted lore. It would seem that
these models could shed light on how we can generalize EFT's beyond our present understanding.

Other recent applications of  EFTs include:
the suggestion that EFTs can be utilized in calculating entanglement  entropy (EE), which has played a central role in quantum information as well as a tool to distinguish phases of matter,
in the limit of  large sub-regions at late times 
via a membrane description.  The EFT description of non-Fermi liquids   based upon discretizing the surface into patches that decouple at low energies leading to membrane-like dynamics with an
enhanced symmetry.

The synergy between the EFT  and the condensed matter community has gained momentum in recent years, with conferences and workshops dedicated to bringing together
physicists from these fields.  Further convergence seems inevitable and will enhance the likelyhood of scientific discovery.

\section{Gravitating systems}

A quantum field theoretic description of gravity must \footnote{Assuming it does not have a sensible UV fixed point.}
be only an effective one, with a cut-off of order the Planck scale. As such, it is possible to
merge gravity and quantum mechanics  in a consistent fashion as long as one is willing to
only study physics at sub-Planckian  scales.  It is perhaps fair to say that the low energy theory of gravity around flat backgrounds is now well understood at the quantum level.
Testing  these quantum predictions are at the moment  beyond the scope of abilities. 
However, recently  EFT's have also  been utilized to
systematize calculations  of classical observables. In particular, the nascent field of gravitational
wave astronomy has become a feeding ground for HEP theorists looking to utilize their field
theoretic skills in a new context~\cite{goldberger}. 

An EFT formalism has been developed to make systematic predictions for binary inspirals
 in the regime where the post-Newtonian (PN)  expansion  is valid, corresponding to the small
 relative velocity limit. This EFT, named NRGR, for Non-Relativistic General Relativity, is a rich
 theory used to factorize the physics arising from all of the relevant length scales  in a binary inspiral including the size of the objects, the orbital radius and the radiation wavelength. 
 The EFT involves integrating out the short distance physics in two stages leading to
 a low energy  theory of a point particle with dynamical radiating multipole moments whose dynamics 
 are dictated by the potential generated when integrating out short distance graviton modes.
 These potentials depend upon Wilson coefficients which encode the internal structure of
 constituents. Extracting the values of these coefficients from the data yields information about
 the equation of state of neutron stars, which depends upon the nature of the QCD ground state
 at large densities.
 
 Using NRGR the community has generated new results for: higher order potentials, multipole moment calculations as well
 a dissipative effects using the in-in closed time path formalism, 
 all of which are necessary ingredients to create gravitational wave templates necessary for
 parameter extraction, the ultimate goal of any gravitational wave measurement.
  Furthermore, it has been shown that
 within the context of the EFT one can utilize scattering amplitudes to stream-line calculations, 
 which has attracted a whole new community to the problem of gravitational wave physics.
 Amplitude methods are especially useful  to study  the Post-Minkowskian  (PM) expansion in powers of $G_N$, 
 which is  systematic when studying near forwards dynamics in the relativistic regime. These fully relativistic calculations, while not systematically applicable to the bound orbit problem, do capture a subset of  the higher order terms in the PN expansion  and  teach us more about the  structure of the near forward limit in GR.

 The use of EFT in gravitational wave physics has spread quite rapidly.  
  There have been numerous dedicated workshops to the EFT of gravitational wave physics and
 the collaborations between HEP and GR scientists is becoming increasingly more common.
 The use of HEP/EFT tools in this field has led to state of the art results. Given the relative newness of this field and the plethora of data that is to come, EFT's will continue to drive theoretical
 predictions. 

\label{sec:gravity}

\section*{Acknowledgements} 
PD acknowledges support from the US Department of Energy under grant number
DE-SC0015655.

\bibliography{TF02_summary_refs}
\bibliographystyle{utphys}

\end{document}